\documentstyle[12pt,a4]{article}

\begin{document}
                     
\bibliographystyle{plain}

\noindent
\centerline{{\large \bf SOLUTION OF THE GAMMA RAY BURST MYSTERY ?}} 

\medskip
\noindent
\centerline{{\bf Nir J. Shaviv and Arnon Dar}}

\medskip
\noindent
\centerline {Department of Physics and Space Research Institute}
\centerline {Israel 
Institute of Technology, Haifa 32000, Israel.}

\bigskip
\centerline {{\bf Abstract}}

\medskip
\noindent
{\bf  
Photoexcitation and ionization of partially ionized heavy atoms in highly 
relativistic flows by interstellar photons, followed by their 
reemission in radiative recombination and decay, boost star-light into 
beamed $\gamma$ rays along the flow direction. 
Repeated excitation/decay of highly relativistic baryonic 
ejecta from merger or accretion induced collapse of neutron stars in 
dense stellar regions (DSRs), like galactic cores, globular clusters and 
super star-clusters, can convert enough kinetic energy in such events 
in distant galaxies into cosmological
gamma ray bursts (GRBs). The model predicts remarkably well all the main 
observed temporal and spectral properties of GRBs. Its   
success strongly suggests that GRBs are $\gamma$ ray tomography pictures 
of DSRs in galaxies at cosmological distances with unprecedented 
resolution: A time resolution 
of $dt\sim 1~ms$ in a GRB can resolve stars at a Hubble distance which 
are separated by only $D\sim 10^{10}cm$. This is equivalent to the  
resolving power of an optical telescope with a diameter larger than 
one Astronomical Unit!}

\vfil
\eject

The origin of gamma ray bursts (GRBs), 
which have been discovered 35 years ago,  
is still a complete mystery [1]. Their observed isotropy in the sky, 
deficiency of faint bursts and the lack of concentration towards the Galactic 
center, in the Galactic disk and in the direction of M31, strongly 
suggest [2] that they are cosmological in origin [3]. A cosmological origin 
implies [4] that GRBs have enormous luminosities during short periods of 
time, \begin{equation}
  L\approx 10^{50}d_{28}^2\chi\phi_6 \Delta\Omega~erg~s^{-1},
\end{equation}
where $d=d_{28} 10^{28}~cm$ is their luminosity distance, 
$\phi=10^{-6}\phi_6~ergs~cm^{-2}s^{-1}$ 
is their measured energy flux, $\chi$ is a bandwidth
correction factor and $\Delta \Omega$ is the solid angle into which their
emission is beamed. Moreover, the short durations of GRBs imply very compact
sources. Relativistic beaming is required then, both in order to explain the
absence of accompanying optical light and X-ray emission, and in order 
to avoid self opaqueness due to $\gamma\gamma\rightarrow e^+e^-$. All 
these considerations  [4] seem to support the favorite cosmological model 
of GRBs; relativistic  fireballs [5] formed by mergers of neutron stars 
(NS) or neutron stars and  black holes (BH) in close binary systems [6] 
due to gravitational wave emission [7], or by accretion induced collapse 
(AIC) of NS and white dwarfs (WD) [6,8]. Indeed, the above considerations 
and the observed rate of GRBs favor NS mergers/AIC as the source of  
cosmological GRBs. Nevertheless, no mechanism has been convincingly 
shown to be able to convert a large enough fraction of their binding 
energy release into $\gamma$ rays and to explain simultaneously the 
complex light curves and spectral behavior of GRBs [4]. 

Most of the binding energy released in NS mergers/AIC is expected to be 
in the form of neutrinos, gravitational waves and kinetic energy of 
ejected material [9]. No mechanism is known which converts efficiently 
gravitational waves into $\gamma$ rays. Neutrino annihilation [6] 
($\nu\bar{\nu} \rightarrow e^+e^-$), and neutrino pair production in 
strong magnetic fields ($\nu\gamma_v\rightarrow \nu e^+e^-$) near 
merging/collapsing NS cannot convert enough binding energy into 
$e^+e^-\gamma$  fireballs or relativistic $e^{\pm}$ beams which can 
produce cosmological GRBs, because of baryon contamination [10] 
and the low efficiency of these processes [11]. Thus, if 
NS-NS and NS-BH mergers, or AIC of WD and NS, 
produce GRBs, the production must proceed through conversion of
baryonic kinetic energy of a highly relativistic flow (a fireball  
or a jet) into $\gamma$ rays. 
Here we propose that repeated
photoexcitation/ionization of the highly relativistic atoms 
of the flow by star-light in dense stellar regions followed by emission of   
decay/recombination photons which are beamed and boosted to 
$\gamma$  ray energies in the observer frame (see Fig. 1), 
produce cosmological GRBs. We show that this simple mechanism, which 
has been overlooked, is able to convert enough baryonic kinetic energy 
of highly relativistic flows in dense stellar regions into cosmological 
GRBs. We also show that it predicts remarkably well the main observed 
properties of GRBs; their burst size, duration distribution, complex 
light curves and spectral evolution.  

We assume that the ejected mass in NS 
mergers or in AIC of NS or WD contains high Z nuclei (e.g., iron-like)
either from the crust of the neutron star or from nucleosynthesis   of ejected 
neutron star matter when it expands and cools [12]. 
In view of the uncertainties in modeling merger/AIC  
of compact stellar objects [11], rather than relying on  
numerical simulations, we deduce the total relativistic kinetic 
energy release in such events from observations of type II supernova 
explosions, which are driven by gravitational core collapse into NS or BH. 
In type II supernova explosions, typically, $10M_\odot$ are accelerated 
to a  final velocity of $v\sim  7000~km~s^{-1}$ [9], i.e., to a total final  
momentum $P\sim 5 M_\odot v$. Since core collapse is not affected 
directly by the surrounding stellar envelope, we assume that a
similar impulse, $\int Fdt\approx P$, is imparted 
to the ejected mass in merger/AIC collapse of NS or WD.  
If the ejected mass is much smaller than a solar 
mass,  $\Delta M \ll M_\odot$, 
then it is accelerated to a highly relativistic velocity. Its kinetic 
energy is given then by $E_K \sim Pc\sim 3\times 10^{53}~erg$. 
About 1\% of this kinetic 
energy must be converted into $\gamma$ rays in order to produce  
cosmological GRBs with $\sim 3\times 10^{51}~erg$. 
 
The natural birth places of close binary systems are the very dense 
stellar regions (DSR) in galactic cores [13], collapsed cores of globular 
clusters (GC) and super star-clusters (SSC) [14]. These DSR are 
rich in very luminous stars and have very large photon  column densities,  
$N_\gamma> L/\pi c R \epsilon~.$ For instance, 
the core of our Milky Way (MW) galaxy has a surface brightness, 
$\Sigma\sim 1.2\times 10^7L_\odot~pc^{-2}$, typical photon energies,  
$\epsilon\sim 1~eV$, and a radius, $R\sim 0.3~pc $ [14]. These values 
yield, on average, $N_\gamma> 10^{23}~cm^{-2}$. 
A similar value is obtained for the nearby M31 galaxy. 
Larger values are possible for the unresolved super star-clusters. 
Much larger values are obtained for active galactic 
nuclei (AGN). The highly relativistic ejecta expands, cools and becomes 
transparent to its own radiation after a relatively short time [5]. 
Its partially ionized, highly relativistic atoms see 
the interstellar photons blue shifted to X-ray energies. These photons  
photoexcite/photoionize the high Z atoms of the flow  
which subsequently decay/recombine radiatively by isotropic X-ray 
emission in the flow's rest frame. For a highly relativistic flow 
with a Lorentz factor $\Gamma \gg 1$, the isotropic X-rays are beamed 
along the flow, and as long as $\theta< 1/\Gamma $ 
their energies, $\epsilon_{X}$, are boosted to $\gamma$ ray 
energies, \begin{equation}
\epsilon_\gamma\approx 2\Gamma \epsilon_{X}/(1+\Gamma^2\theta^2),
\end{equation} 
where $\theta$ is the angle of the flow direction relative to the 
observer. The differential fraction of the emission that is directed 
towards the observer  is given by 
\begin{equation}
dI/d(cos\theta) \approx 2\Gamma^2/(1+\Gamma^2\theta^2)^2.
\end{equation} 
 
The total baryonic kinetic energy which is converted into 
beamed $\gamma$ ray emission by repeated excitation/decay  
can be estimated as follows: The total photoabsorption cross section of 
an atom in an atomic state $n$ into a state $n'$ is 
\begin{equation}
  \int \sigma_{\nu nn'}d\nu={\pi e^2\over m_ec}f_{nn'},
\end{equation}
where the integral is over the natural line width and where the oscillator
strength $f_{nn'}$ satisfies the sum rule, $\sum_{n'} f_{nn'}=Z$, with Z being
the number of electrons in the system [15]. If the typical energy of the
interstellar photons is $\epsilon \sim 1~eV$, and if their energy is 
boosted in the flow's rest frame by $\Gamma\sim 10^3$, 
then their typical photoabsorption cross section by Fe atoms is
\begin{equation}
 \bar{\sigma}\equiv {\int \sigma_{\nu nn'}hd\nu\over \Gamma \epsilon}
={\alpha\over 2} {h\over m_ec}{hcZ\over\Gamma \epsilon}
 \approx 3\times 10^{-18}~cm^2.
\end{equation}
This cross section is larger by about seven orders of magnitude 
than the typical cross sections for production of $\gamma$ rays 
in the interstellar medium by 
inverse Compton scattering, bremstrhalung and synchrotron emission by 
electrons.
Thus, if the ionized or excited atoms recombine/decay radiatively 
fast enough, then
the typical energy of GRBs (burst size) from NS mergers/AIC  
in GC is given approximately by
\begin{equation}
 E_\gamma \approx \bar{\sigma}N_\gamma(\Gamma \epsilon/ m_{A}c^2)  
 E_K \approx 3\times 10^{51}~erg,        
\end{equation} 
where $m_{A}\approx 56m_p$. 
In its rest frame the relativistic 
flow expands against the external radiation field until its 
internal pressure equals the external pressure due to
photoabsorption of the Lorentz boosted photons of the radiation field.
Its relativistic atoms are ionized by the photoabsorption of the 
Lorentz boosted interstellar photons and than recombine radiatively and 
decay  by line emission. The ionization/excitation rate adjusts itself 
to the recombination/decay rate. A detailed treatment [16] shows that the 
relativistic flow cools to rather a low temperature (eV), its Fe atoms 
are only partially ionized and it has a relatively large density which 
indeed result in fast enough radiative recombination and decay , 
except, perhaps, near AGN. However, if NS merger/AIC does occur near a 
bright AGN, where $L\sim 10^{12}L_\odot$ and $R\sim 10^{17}~cm $, 
i.e., $N_\gamma\sim 10^{29}~ cm^{-2}$, then a large enough fraction of 
the kinetic energy of the relativistic flow can be converted into a GRB 
even by inverse Compton scattering [17,18].  Such a GRB will  
be extremely short and structureless.

To calculate the detailed predictions of our model, we have constructed 
a numerical Monte Carlo code which simulates the formation of cosmological 
GRBs by highly relativistic flows in dense stellar regions [16]. The code 
employs the quantum mechanical cross sections for the relevant photo 
excitation/ionization and radiative recombination/decay processes. 
The core of the MW galaxy [13] was  used for modeling the stellar 
environment (density of stars, stellar luminosities and stellar 
temperatures) in a typical GC. Initial distributions of $\Gamma$   
which are consistent with theoretical considerations (e.g., shock 
acceleration) and observations have been used. Using the GRB simulation 
code we have found [16] that (a) the main properties of the simulated 
GRBs are not sensitive to fine details and (b) the calculated light curves 
and spectral behavior of simulated GRBs reproduce remarkably well those 
observed in GRBs [1, 19-21]. In particular, the simulated GRBs look
undistinguishable from the observed GRBs. This is demonstrated in Fig. 
2 which presents a simulated light curve of a
GRB and its power spectrum, $P(w)\equiv \vert\int L(t)exp(iwt)dt\vert^2$,  
that has the universal power-law 
behavior $P(w)\sim  w^{-2}$, which was found for real GRBs [18].
This simple universal behavior can be 
derived analytically [16,18] from our model. Also the other 
main observed properties of GRBs [1,19-21] follow from our model. Here we 
summarize briefly their approximate  analytical derivations [16].  
For the sake of simplicity we neglect here general relativistic effects 
(e.g., time dilation and energy redshift ) and assume that 
the explosions are spherical symmetric and occur at the center of the GCs, 
that the stars within a GC have a uniform spatial 
distribution, the same luminosity and the same effective surface  
temperature, and that the energy flux in the relativistic flow, 
$E^2dn_{A}/dE$, is peaked around a Lorentz 
factor $\Gamma$.  Then our model predicts that:

1. GRB light curves are composed of a smooth background plus 
strong and weak pulses. Strong pulses are  produced when the 
relativistic flow passes 
near luminous stars within the ``beaming cone'', i.e., stars at an angle 
$\theta_*< 1/\Gamma$ relative to the line of sight from the explosion 
to the observer (see eq. 2). 
A multiple star system (binary, triplet, etc) produces  
a multipeak pulse with very short time-spacing 
between the peaks (spikes). Weak pulses are produced by boosting
star-light from less luminous stars or by luminous stars outside the beaming 
cone. The smooth background is produced by boosting the background light 
in the beaming cone from all the other stars in the GC. 

2. The duration of a GRB reflects the spread in arrival
times of gamma rays produced within the beaming cone 
(we neglect the short formation time of the relativistic flow because the 
dynamical time for NS merger/AIC is much shorter, typically few ms). It 
is given approximately by 
\begin{equation}
T\sim  R/c\Gamma^2.
\end{equation}
Hence, relativistic flows with $\Gamma\sim 10^3$ in MW-like
GC ($R\sim 0.3~pc$) produce GRBs that last typically  30 seconds. 
Thus, approximately a one light-year path in a GC is  
contracted into 30 seconds $\gamma$ ray picture in the observer 
frame.     
 
3. The properties of the pulses from single stars
depend on their locations, luminosities and 
spectra, and on the distribution of $\Gamma$ in the 
flow.  A pulse from a star at a distance $D_*$  and an angle 
$\theta_*$ begins at a time $t_i\sim D_*\theta_*^2/2c$
after the beginning of the GRB ($t\equiv 0)$. 
Its duration (integrated intensity equals that due to the whole GC  
background) is given approximately by 
\begin{equation}
T_p\sim D_*\theta_*d\theta_*\approx b/c\Gamma \sim (4\pi/3) R L_*/\pi 
c\Gamma L, \end{equation}
where the integrated intensity from boosting photons from the star over 
impact parameters smaller than b equals that due to all the other stars in  
the GC. For main sequence stars with $10^{-1}\leq L_*/L_\odot\leq 10^3,$ 
$\Gamma=10^3$ and MW-like GC we obtain $10~ms\leq T_p\leq 100~s$.
For typical early type stars with $L_*\sim 3-6 L_\odot$, $T_p\sim 0.3-0.6~ 
s$. Multiple stars yield multipeak pulses. For instance,
the time difference between the two pulses from a binary star is given by
\begin{equation}
T_b\approx D_*\theta_*d\theta_*/c\approx d_p/\Gamma c, 
\end{equation}
where $d_p$ is the distance between the binary stars projected on the
plane perpendicular to the flow. Thus, $T_b<500(d_{\rm A.U.}
/\Gamma_3)~ms,$ where $d_{\rm A.U.}$ is the binary separation in 
astronomical units. Since a large fraction of the stars are in close binaries,  
triplets, etc, a large fraction of the pulses have a multipeak 
structure. 
 
4. In a GRB from a GC with $N_*$ stars uniformly distributed,    
the average number of stars within the beaming cone 
is $n_p\sim N_*/4\Gamma^2$. Thus, for a MW-like GC and $\Gamma\sim 10^3$,
we expect, on average, $n_p\sim 3$ stellar pulses. Their rate 
is $dn_p/dt\sim cN_*(r)/4r\sim 0.1~s^{-1}.$ It is independent of time if
$N_*(r)$, the number of stars within a radius r, is proportional to r,
as seen in the GC of the MW [13].    
Because $n_p$ is a small number, $n_p$ and the average 
time-spacing between strong pulses, $\Delta T$, which is given by 
\begin{equation}
\Delta T=T/ n_p\sim 4R/cN_*\sim 10 s,
\end{equation}
are predicted to fluctuate considerably around their average values.

5. The total number of gamma rays in a stellar pulse arriving 
from the explosion is given by 
\begin{equation}
\int F_\gamma dt\approx 
10^3\sigma_{17}(L_*/L_\odot)/d_{28}^2\epsilon_{ev}\Gamma_3^2t_{10}~~m^{-2}     
\end{equation}
where $t_*=D_*/c\Gamma^2=10t_{10}~s$ is the peak intensity time of 
the stellar pulse and $\sigma=10^{-17}\sigma_{17}~cm^2$.

6. The durations of GRBs have a bimodal distribution which is a trivial 
consequence of their multipulse nature and the fact that, on 
average, $T_p\ll\Delta T$ independent of $\Gamma$: Multipulse GRBs 
have durations which are equal approximately to the sum of the  
time-spacing between their pulses. 
Consequently, $T\sim \Sigma\Delta T\gg T_p$. This produces 
a bimodal distribution which peaks around 
$T_p\sim 0.3~s$ for single pulse GRBs (see point 3) and 
around $T\sim 30~s$ for multipulse GRBs. 
This is demonstrated in Fig. 3 for simulated explosions in a MW-like GC. 
 
7. Although the photoexcitation 
of partially ionized atoms produces line emission, the line emission is  
Doppler broadened into a 
continuum by the continuous distribution of $\Gamma$ of the 
atoms in the flow. Thus, the model predicts a continuous energy spectrum. 
The minimal Lorentz factor in the flow, $\Gamma_m$, and the
ionization state of the atoms cuts off 
emission in the observer frame below  $E_{min}\sim \Gamma_m{\rm I}\sim 
5-25~keV $, where ${\rm I\sim few\times 100~eV }$ is the ionization 
potential of the last bound electrons in the partially ionized Fe atoms 
[16]. It explains why GRBs are not accompanied by detectable X-ray or 
optical-light emission.   
 
8. If the spectrum of $\Gamma$  in the flow
has a power-law form, $dn_{A}/d\Gamma\sim \Gamma^{-p}$ for 
$\Gamma>\Gamma_m$, then  
the energy spectrum of a strong pulse is given approximately by 
\begin{equation}
dn_\gamma/dE \sim \bar{\sigma}(\Gamma)n_{eff}(\Gamma)(dE/d\Gamma)^{-1}
\Gamma^{-(p-1)}/(\Gamma\theta_*+1/2)
\end{equation}
where $\bar{\sigma}
\sim \Gamma^{-(1+\delta)}$ and $n_{eff}$ is the average number of emitted 
photons per excitation/ionization. 
For $\Gamma\gg 1/\theta_*$, one has [16] $n_{eff}\propto \Gamma$,
$E\sim 2\Gamma \epsilon_{X}$,  $\delta\leq 1$ and then 
$ dn_\gamma/dE\sim E^{-(p+\delta)}.$ 
For  $\Gamma\ll 1/ \theta_*$, one has [16] $n_{eff}\approx 1$,     
$E\sim \epsilon\Gamma^2$, $\delta\sim 0$, and then 
$ dn_\gamma/dE\sim E^{-(p+1)/2}.$ Hence the   
spectrum of a pulse has a broken power-law form. The break occurs when  
$2\Gamma\theta_*\sim 1$, i.e., around an 
energy $E_b\approx 2\Gamma\epsilon_{X}\approx \epsilon_{X}/\theta_*~,$
and the power-index changes by $\sim (p+1)/2\sim 1.5$. 
Since  $t_i\sim D_*\theta_*^2/2c$ and $D_*\approx R$, on 
average, $E_b\sim 1/t_i^\alpha $ with $1/2<\alpha<1$. 
(For a GC with a uniform stellar distribution, the probability of 
$D_*$ is proportional to $D_*^2$ and consequently 
most of the stars have $D_*\approx R$.) 

9. The relative arrival times at a star of atoms with different  
$\Gamma$ are given by  $t-t_i\approx D_*/2c\Gamma^{-2}$. Such 
atoms boost the star-light to an energy between $\epsilon _\gamma\sim 
\epsilon\Gamma^2$ and $\epsilon _\gamma\sim 2 \epsilon_{X}\Gamma$.     
Therefore, the peak energy, 
$E_p\equiv max ~E^2(dn_\gamma/dE)$, decreases during a pulse   
approximately as $\sim 1/(t-t_i+\delta t_i)^\alpha~$ i.e.,    
\begin{equation} 
E_p\sim \Sigma_i A_i t_i^{-\alpha}(t-t_i+\delta t_i)^{-\alpha}~; 1/2\leq 
\alpha\leq 1, 
\end{equation}  
where $A_i$ is proportional to the mean photon energy of the star-light and 
$\delta t_i $ is an added time broadening (a few ms) due to 
the time extension of the explosion and the extension of the 
region around the star where the boosting of the star-light   
takes place. $E_p$ is maximal right after the 
pulse begins and decreases monotonicly afterwards while the photon flux, 
$dn_\gamma/dE$, peaks at a later time, reflecting the $1/r^2$ behavior
of the intensity of the radiation around a star and the
distribution of $\Gamma$ in the flow. Normal power-law 
spectra  yield $\gamma$ ray fluxes during pulses that, 
on average, are time-asymmetric (fast rise and a slower decay) 
with longer pulses being more asymmetric.
Pulses are narrower and their 
peak luminosities are shifted closer to the beginning of the pulse when 
viewed in higher energy bands (larger $\Gamma$) as demonstrated in Fig. 4.
  
10. For a MW-like GC, $N_\gamma$ is not large enough to attenuate the
relativistic flow. The kinetic energy 
of the the ejecta is injected into the galactic and intergalactic space. 
The general consequences of this injection will be discussed elsewhere. 
Here,  we comment only on the high energy $\gamma$ ray production due to 
occasional 
collision of the ejecta with interstellar matter (stellar winds, planetary 
nebulae, etc) or with a molecular cloud with a sizable column density 
along the line of sight, in/near the GC (10\% of the solid angle around AGN 
is covered by hydrogen clouds). Typical clouds, 
have  $R\sim 5 ~pc$ and $M\sim 10^4M_\odot$, yielding proton column 
densities of $N_p\sim 10^{22}cm^{-2}$. The total inelastic cross 
section of high energy iron nuclei on protons is $\sigma\approx 
10^{-24}cm^{-2}$. Thus, nuclear collisions of the flow   
with a molecular cloud near the explosion 
will produce pions, and consequently, a burst of $\sim 10^{51}erg$
multi GeV $\gamma$ rays through $\pi^0\rightarrow 2\gamma$ decays
(and neutrinos through $\pi^{\pm}\rightarrow \mu\nu_\mu$ and 
$\mu\rightarrow e\nu_e\nu_\mu$ decays), with a power-law
spectrum, $dn_\gamma/dE\sim E^{-p}$. For $p\sim 2$, production of $\sim 20~
GeV$ $\gamma$ rays comes from $\pi^0$'s  produced mainly by nuclei 
with $\Gamma\sim 200$. Thus $\pi^0$ produced $\gamma$ rays of $20~GeV$ 
are delayed, typically, by $R/2c\Gamma^2\sim 2h$,    
as was observed in the case of the 17 February 1994 GRB [22].  

In conclusion, we have described a simple mechanism by which
NS mergers/accretion induced collapse in dense galactic cores 
produce cosmological GRBs. 
The remarkable success of the model in 
reproducing all the main observed temporal and spectral properties of GRBs  
[16] strongly suggests that GRBs are $\gamma$ ray tomography pictures of 
DSR of distant galaxies with unprecedented resolution: A time resolution
of $dt\sim 1~ms$ in GRBs can resolve stars at a Hubble distance of 
$3000h^{-1}Mpc \approx 10^{28}h^{-1}cm$ that are separated by only  
$\sim c\Gamma dt\sim 3\times 10^{10}cm$. This is equivalent to the  
resolving power, $\sim \lambda/D$, of an optical telescope
with a diameter $D>2\times 10^{13} cm$, i.e. more than one 
Astronomical Unit!  

Finally, the proposed mechanism may play
an important role also in other astrophysical gamma ray sources such 
as AGN, pulsars and other cosmic accelerators. 

{\bf Acknowledgement}: We thank G.J. Fishman, A. Laor, 
J.P. Lasota, L. Okun and A. Ori for useful comments and 
the Technion fund for the promotion of research for support.
This research was not supported by the Israel Science Foundation. 
\vfil
\eject
     \parindent 0cm
\centerline{References}

[1] See, e.g., G.J. Fishman and C.A.A. Meegan, Ann. Rev. Astr. 
 Ap. {\bf 33}, 415 (1995).  

[2] See, e.g., M.S. Briggs, 
Ap. \& Sp. Sc.  {\bf 231}, 3 (1995).

[3] V.V. Usov, and G.B. Chibisov, Sov. Astr. {\bf 19}, 115 (1975);
S. van den Bergh, Astr. \& Ap. Suppl. {\bf 97}, 385 (1983).

[4] For a review of cosmological models see C.D. Dermer 
and T.J. Weiler,  Ap. \& Sp. Sc.  {\bf 231}, 377 (1995).

[5] G. Cavallo and M.J. Rees, Mon. Not. Roy. Astr. Soc. {\bf 183}, 359
(1978); B. Paczynski, Ap. J. {\bf 308}, L43 (1986); J. Goodman, Ap. J. 
{\bf 308}, L47 (1986).

[6] J. Goodman, A. Dar and S. Nussinov,  Ap. J. {\bf 314}, L7 (1987);
D. Eichler et al., Nature, {\bf 340}, 126 (1989).

[7] R.A. Hulse and J. H. Taylor, Ap. J., {\bf 368}, 504 (1975).

[8] A. Dar et al., Ap. J. {\bf 388}, 164 (1992); S.E. Woosley, Ap. J. 
{\bf 405}, 273 (1992).

[9] See, e.g., S. L. Shapiro and S.A. Teukolsky,  ``Black 
Holes, White Dwarfs and Neutron Stars'' (A. Wiley Intersc. Pub.) 1983.

[10] B. Paczynski, Ap. J. {\bf 363}, 218 (1990).

[11] H.T. Janka and M. Ruffert, Astr. \& Ap. {\bf 307},
L33 (1996).  

[12] G.J. Fishman, private communication.

[13] D.A. Allen, ``The Nuclei of Normal Galaxies'' (eds. R. Genzel \& 
A.I. Harris, 1994) p. 293; P. Saha et al., Ap. J. in press (1996), 
astro-ph/9604153.      

[14] D. Maoz et al., Ap. J.. in press (1996). 

[15] See, e.g., Ya. B Zel'dovich and Yu. P. Raizer, 
``Physics of
Shock Waves and High Temperature Hydrodynamic Phenomena'' (Academic Press 
1967).

[16] For details see N.J. Shaviv, Ph.D Thesis, 1996 (Technion Report 
Ph-96-16).

[17] A. Shemi, Mon. Not. Roy. Astr. Soc. {\bf 269}, 1112 (1994). 

[18] N.J. Shaviv and A. Dar, Mon. Not. Roy. Astr. Soc. {\bf 277}, 287 
(1995). 
 
[19] See, e.g., J.P. Norris, et al., Ap. J. {\bf 459}, 333 
(1996) and references therein.
 
[20] See, e.g.,  D.L. Band et al., Ap. J. {\bf 413}, 231 (1993); L.A. Ford
et al., Ap. J. {\bf 439}, 307 (1995); B.J. Teegarden , Astr. \& Sp. Sc. {\bf 
231}, 137 (1995).

[21] G.J. Fishman et al., Ap. J. Suppl. {\bf 92}, 229 (1994).
 
[22] K. Hurley et al., Nature {\bf 372}, 652 (1994).

\centerline{Figure Captions}

Fig.1  A schematic drawing illustrating the formation of a GRB
by a highly relativistic spherical flow in a dense stellar region. Most of
the observed $\gamma$ rays are produced by the radiative decay of
photoexcited atoms near stars within a cone with an opening angle 
$\theta\sim 1/\Gamma$ along the direction to the observer. 

Fig.2  A typical light curve of a simulated GRB and its temporal power 
spectrum. The straight line represents a $w^{-2}$ power law 
behavior. 

Fig.3  A   duration distribution of simulated GRBs from a MW-like GC.

Fig.4  A simulated GRB from a MW-like GC viewed in different energy bands.

\end{document}